\documentclass{aa}

\usepackage[T1]{fontenc}
\usepackage{pslatex}
\usepackage{amsmath}
\usepackage{amsfonts}
\usepackage{xcolor}
\usepackage{url}
\usepackage{bm}
\usepackage{graphicx}
\usepackage{amssymb}
\usepackage{soul}
\usepackage{booktabs}
\usepackage{array}
\usepackage[utf8]{inputenc}
\inputencoding{latin1}
\inputencoding{utf8}
\usepackage{appendix}

\def\lsim{ \lower .75ex\hbox{$\sim$} \llap{\raise .27ex \hbox{$<$}} }
\def\gsim{ \lower .75ex \hbox{$\sim$} \llap{\raise .27ex \hbox{$>$}} }

\newcommand{\bi}{\begin{itemize}}
\newcommand{\ei}{\end{itemize}}

\newcommand{\ixpe}{{IXPE}\xspace }
\newcommand{\ixpeobssim}{{\texttt{ixpeobssim}}\xspace}
\newcommand{\pcube}{{\texttt{PCUBE}}\xspace}
\newcommand{\xspec}{{\texttt{XSPEC}}\xspace}
\newcommand{\const}{{\texttt{constant}}\xspace}
\newcommand{\tbab}{{\texttt{tbabs}}\xspace}
\newcommand{\polconst}{{\texttt{polconst}}\xspace}

\newcommand{\logpar}{{\texttt{logpar}}\xspace}
\newcommand{\I}{{\it I}\xspace}
\newcommand{\Q}{{\it Q}\xspace}
\newcommand{\U}{{\it U}\xspace}

\newcommand{\pdx}{{$\Pi_{\rm X}$}\xspace}
\newcommand{\pax}{{$\psi_{\rm X}$}\xspace}
\newcommand{\ergsc}{{$\rm{erg\;s^{-1}\;cm^{-2}}$}\xspace}

\usepackage[colorlinks=true,linkcolor=blue,citecolor=blue]{hyperref}

\begin{document}

\title{The {\it IXPE}  and multifrequency polarimetric view of the extreme blazars 1ES 1101-232 and RGB J0710+591}
\titlerunning{Multifrequency polarimetry of 1ES 1101-232 and RGB J0710+591}

\author{
Fabrizio Tavecchio\inst{\ref*{merate}}\orcid{0000-0003-0256-0995} \and 
Dawoon E. Kim\inst{\ref*{iaps}}\orcid{0000-0001-5717-3736} \and   
Gabriel Emery\inst{\ref*{granada}}\orcid{0000-0001-6155-4742} \and 
Ioannis Liodakis \inst{\ref*{forth}}\orcid{0000-0001-9200-4006} \and
Iv\'an Agudo\inst{\ref*{granada}}\orcid{0000-0002-3777-6182} \and
Paolo Coppi\inst{\ref*{yale}}\orcid{0000-0001-9604-2325} \and 
Giampiero Tagliaferri\inst{\ref*{merate}}\orcid{0000-0003-0121-0723} \and Laura Di Gesu\inst{\ref*{asi}}\orcid{0000-0002-5614-5028}  \and 
Tullia Sbarrato\inst{\ref*{merate}}\orcid{0000-0002-3069-9399}   
\and Lucia Ballo\inst{\ref*{esa}}\orcid{0000-0002-5036-3497}\and 
Alberto Sciaccaluga\inst{\ref*{merate}}\orcid{0000-0001-6181-839X}\and 
Steven R. Ehlert\inst{\ref*{marshall}}\orcid{0000-0003-4420-2838}  \and
  Giacomo Bonnoli\inst{\ref*{merate}} \orcid{0000-0003-2464-9077} \and 
  Francisco Jos\'e Aceituno\inst{\ref*{granada}} \and  
Carolina Casadio\inst{\ref*{forth},\ref*{crete}}
  V\'{i}ctor Casanova\inst{\ref*{granada}} \and 
 Immacolata Donnarumma\inst{\ref*{asi}}\orcid{0000-0002-4700-4549} \and
  Juan Escudero\inst{\ref*{harv_smithsonian} \and \ref*{granada}}\orcid{0000-0002-4131-655X} \and 
  Daniel Morcuende\inst{\ref*{granada}} \and 
  Jorge Otero-Santos\inst{\ref*{padova},\ref*{granada}} \and 
  Alfredo Sota\inst{\ref*{granada}} \and 
  Vilppu Piirola\inst{\ref*{physics_finland}} \and 
  Pouya M. Kouch\inst{\ref*{physics_finland},\ref*{finnish_eso}}\orcid{0000-0002-9328-2750} \and 
  Elina Lindfors\inst{\ref*{physics_finland}}\orcid{0000-0002-9155-6199} \and 
  Kari Nilsson\inst{\ref*{finnish_eso}}\orcid{0000-0002-1445-8683} \and 
  Ioannis Myserlis\inst{\ref*{mm_granada}}\orcid{0000-0003-3025-9497} \and 
  Mark Gurwell\inst{\ref*{harv_smithsonian}}\orcid{0000-0003-0685-3621} \and 
  Garrett Keating\inst{\ref*{harv_smithsonian}}\orcid{0000-0002-3490-146X} \and 
  Ramprasad Rao\inst{\ref*{harv_smithsonian}}\orcid{0000-0002-1407-7944} \and 
  Emmanouil Angelakis\inst{\ref*{bonn_address}}\orcid{0000-0001-7327-5441} \and 
  Alexander Kraus\inst{\ref*{planck_bonn}}\orcid{0000-0002-4184-9372} \and 
  Ryo Imazawa\inst{\ref*{physics_hiroshima}} \and 
  Mahito Sasada\inst{\ref*{physics_tokyo}} \and 
  Yasushi Fukazawa\inst{\ref*{physics_hiroshima},\ref*{astro_hiroshima},\ref*{core_hiroshima}} \and 
  Koji S. Kawabata\inst{\ref*{physics_hiroshima},\ref*{astro_hiroshima},\ref*{core_hiroshima}} \and 
  Makoto Uemura\inst{\ref*{physics_hiroshima},\ref*{astro_hiroshima},\ref*{core_hiroshima}} \and 
  Tsunefumi Mizuno\inst{\ref*{astro_hiroshima}}\orcid{0000-0001-7263-0296} \and 
  Tatsuya Nakaoka\inst{\ref*{astro_hiroshima}} \and 
  Sumie Tochihara\inst{\ref*{physics_hiroshima}} \and 
  Takahiro Akai\inst{\ref*{physics_hiroshima}} \and 
  Hiroshi Akitaya\inst{\ref*{chiba_jap}} \and 
  Rumen Bachev\inst{\ref*{bulgaria}} \and 
  Anton Strigachev\inst{\ref*{bulgaria}} \and 
  Petra Benke\inst{\ref*{geo_potsdam}, \ref*{planck_bonn}} \and 
  Lena Debbrecht\inst{\ref*{planck_bonn}} \and 
  Julia Eich\inst{\ref*{wurzburg}} \and 
  Florian Eppel\inst{\ref*{planck_bonn}, \ref*{wurzburg}} \and 
  Andrea Gokus\inst{\ref*{physics_washington}} \and 
  Steven H\"{a}mmerich\inst{\ref*{remeis_obs}} \and 
  Jonas He\ss d\"orfer\inst{\ref*{planck_bonn}, \ref*{wurzburg}} \and 
  Matthias Kadler\inst{\ref*{wurzburg}} \and 
  Sanghyun Kim \inst{\ref*{KASI}, \ref*{U_Korea}} \and 
  Dana Kirchner\inst{\ref*{wurzburg}} \and 
  Georgios Filippos Paraschos\inst{\ref*{planck_bonn}}\orcid{0000-0001-6757-3098} \and 
  Florian R\"{o}sch\inst{\ref*{planck_bonn}, \ref*{wurzburg}} \and 
  Wladislaw Schulga\inst{\ref*{wurzburg}} 
}
\authorrunning{Tavecchio et al.}

\institute{
INAF -- Osservatorio Astronomico di Brera, Via E. Bianchi 46, I-23807 Merate, Italy \label{merate}
\email{fabrizio.tavecchio@inaf.it}
\and
INAF -- Istituto di Astrofisica e Planetologia Spaziali, Via Fosso del
Cavaliere, 100 - I-00133 Rome, Italy \label{iaps}
\and
  Instituto de Astrof\'{i}sica de Andaluc\'{i}a, IAA-CSIC, Glorieta de la Astronom\'{i}a s/n, 18008 Granada, Spain \label{granada}
\and
Institute of Astrophysics, Foundation for Research and Technology-Hellas, Vasilika Vouton, GR-70013 Heraklion, Greece\label{forth}
\and
Department of Astronomy, Yale University, PO Box 208101, New Haven, CT 06520-8101, USA \label{yale}
\and
ASI - Agenzia Spaziale Italiana, Via del Politecnico snc, 00133 Roma, Italy \label{asi}
\and
European Space Astronomy Centre (ESA/ESAC), 28691 Villanueva de la Canada Madrid, Spain \label{esa}
\and
NASA Marshall Space Flight Center, Huntsville, AL 35812, USA \label{marshall}  
\and
  Department of Physics, University of Crete, Voutes University Campus, 70013 Heraklion, Greece  \label{crete} 
\and
  Center for Astrophysics | Harvard \& Smithsonian, 60 Garden Street, Cambridge, MA 02138 USA \label{harv_smithsonian} \and
  Istituto Nazionale di Fisica Nucleare, Sezione di Padova, 35131 Padova, Italy \label{padova} \and
  Department of Physics and Astronomy, 20014 University of Turku, Finland \label{physics_finland} \and
  Finnish Centre for Astronomy with ESO, 20014 University of Turku, Finland \label{finnish_eso} \and
  Instituto de Radioastronomía Millimétrica, Avenida Divina Pastora, 7, Local 20, E–18012 Granada, Spain \label{mm_granada} \and
  Orchideenweg 8, 53123 Bonn, Germany \label{bonn_address} \and
  Max-Planck-Institut f\"{u}r Radioastronomie, Auf dem H\"{u}gel 69, D-53121 Bonn, Germany \label{planck_bonn} \and
  Department of Physics, Graduate School of Advanced Science and Engineering, Hiroshima University Kagamiyama, 1-3-1 Higashi-Hiroshima, Hiroshima 739-8526, Japan \label{physics_hiroshima} \and
  Department of Physics, Tokyo Institute of Technology, 2-12-1 Ookayama, Meguro-ku, Tokyo 152-8551, Japan \label{physics_tokyo} \and
  Hiroshima Astrophysical Science Center, Hiroshima University 1-3-1 Kagamiyama, Higashi-Hiroshima, Hiroshima 739-8526, Japan \label{astro_hiroshima} \and
  Core Research for Energetic Universe (Core-U), Hiroshima University, 1-3-1 Kagamiyama, Higashi-Hiroshima, Hiroshima 739-8526, Japan \label{core_hiroshima} \and
  Planetary Exploration Research Center, Chiba Institute of Technology 2-17-1 Tsudanuma, Narashino, Chiba 275-0016, Japan \label{chiba_jap} \and
  Institute of Astronomy and NAO, Bulgarian Academy of Sciences, 1784 Sofia, Bulgaria \label{bulgaria} \and
  GFZ Helmholtz Centre for Geosciences, Telegrafenberg, 14476, Potsdam, Germany \label{geo_potsdam} \and
  Julius-Maximilians-Universit\"{a}t W\"{u}rzburg, Institut f\"{u}r Theoretische Physik und Astrophysik, Lehrstuhl f\"{u}r Astronomie, Emil-Fischer-Stra{\ss}e 31, 97074 W\"{u}rzburg, Germany \label{wurzburg} \and
  Physics Department and McDonnell Center for the Space Sciences, Washington University in St. Louis, MO, 63130, USA \label{physics_washington} \and
  Dr. Karl-Remeis Sternwarte and Erlangen Centre for Astroparticle Physics, Friedrich-Alexander Universit\"at Erlangen-N\"urnberg, Sternwartstr.~7, 96049 Bamberg, Germany \label{remeis_obs} \and
  Korea Astronomy and Space Science Institute, 776 Daedeok-daero, Yuseong-gu, Daejeon 34055, Korea \label{KASI} \and
  University of Science and Technology, Korea, 217 Gajeong-ro, Yuseong-gu, Daejeon 34113, Korea \label{U_Korea}
}
\date{}

\clearpage

\abstract{Multiwavelength polarimetry is a powerful tool to probe magnetic field and flow geometries in the relativistic jets of blazars. In this respect, particularly interesting are the sources whose synchrotron emission covers a broad range of frequencies, from radio to X-rays, such as the BL Lac objects of the HSP type. Previous measurements including radio, optical and X-ray data show a clear trend, with the degree of polarization increasing with frequency. 
Here we report radio, optical and X-ray observations ($Swift$, $Nustar$ and $IXPE$) of 1ES 1101-232 and RGB J0710+591, two blazars belonging to the puzzling subclass of extreme BL Lacs (EHBL). For 1ES 1101-232 we found a strong frequency-dependency of the degree of polarization (with a ratio $\Pi_X/\Pi_O\simeq 5.2$). For RGB J0710+591, IXPE derived a 1$\sigma$ upper limit $\Pi_X<11.6\%$, comparable with the measured optical degree of polarization (average $\Pi_O\sim 12\%$). We discuss the results in the framework of current interpretations and, in particular, we report an improved version of the stratified shock model that is able to reproduce the observed data of both sources. 
}


\maketitle

\section{Introduction}

Despite decades of observational and theoretical efforts, the physical processes acting in relativistic jets are still in part unknown. Basic questions concerning, e.g., the matter content of the jet, the role of the magnetic field, and the emission mechanisms are still open \citep{blandford19}. One of the most central topics concerns the mechanisms able to energize the emitting particles at the required ultra-relativistic energies. In the past, shocks were identified as the most likely acceleration sites \citep[e.g.][]{blandford87}, but recent studies highlighted the potential role of magnetic reconnection and turbulence \citep[e.g.][]{matthews20}. 

Due to the relativistic amplification of the jet emission caused by favorable alignment, the physical processes at work in the jet near the central engine are best studied in blazars \citep{blandford19,romero17}. The emission from this class of sources spans the entire electromagnetic spectrum, from radio waves up to high-energy $\gamma$ rays \citep{fossati98}. The ``double hump" shape of the spectral energy distribution (SED) flags the presence of two emission components. The low energy bump (peaking, depending on the source type, from IR to the X-ray band) is produced by electrons (produced at distances of the order of 100-1000 gravitational radii from the central black hole) through synchrotron emission. The origin of the high-energy component is still debated, with leptonic and hadronic models as contenders \citep[e.g.][]{cerruti20,SolZech22}. 
It is worth noticing that most of the models used to reproduce the emission of blazars (both leptonic and hadronic) adopt a one-zone scenario, which assumes that most of the radiation is produced by particles contained within a unique region of the jet characterized by a homogeneous physical structure.

High synchrotron peaked (HSP) blazars, where the synchrotron component peaks in the X-ray band and the high-energy component reaches TeV energies, are particularly interesting for studies of particle acceleration mechanism(s).  Their intense X-ray emission allows one to probe in detail the dynamics of the emitting particles, and to track the main processes at work (e.g. acceleration, cooling) at the rapidly-evolving high energy tail of the electron energy distribution. Indeed, due to the short cooling length of electrons emitting at these energies, their emission can be used to probe the regions in the vicinity of the acceleration site. In particular, it has been anticipated that the properties of the polarization in this band could be exploited to discriminate among the potential acceleration mechanisms at work \citep{tavecchio21}. Indeed, the highly successful {\it IXPE} observations of a handful of HSP strongly support shock as the main actor in the acceleration process \citep{liodakis22} and make possible for the first time a physical description of the emission region(s)\footnote{ While obtained for a specific class of sources, these results are likely valid also for other kind of astrophysical objects hosting relativistic jets}. In particular, the common trend displayed by HSP is a strong frequency-dependent fraction of polarization (see e.g. \citealt{Kim24}), expected by the "stratified shock'' scenario \citep{angelakis16,tavecchio18}, where, contrary to the widely adopted one-zone scheme, electrons accelerated at the shock cool within an energy-dependent distance of the downstream flow, thus experiencing a different degree of order of the magnetic field, eventually encoded in the polarization (but see \citealt{Bolis24} for an alternative view).

Among HSP, the group of {\it extreme HSP} (a.k.a. EHBL) \citep{costamante01} is attracting growing attention since their unusual properties are difficult to reproduce in the framework of standard models \citep{Katar06,Bonnoli15,costamante18,biteau20}. At odds with the bulk of the blazar population, their SED is generally stable, showing only weak (factor of 2) variability over long timescales (weeks). Moreover, their extremely hard $\gamma$-ray continuum (locating the peak of the high-energy component above 10 TeV) is extremely difficult to interpret in standard models based on inverse Compton emission (e.g., \citealt{biteau20}). The unusual phenomenology of EHBL has been explained invoking a hard electron distribution with a large minimum energy \citep{Katar06,tavecchio09}, a maxwellian-like electron distribution \citep{Lefa11}, internal absorption \citep{aharonian08} or emission from a large-scale jet \citep{boettcher08}.
The peculiarities of EHBL have recently been interpreted as the result of a special shape for the energy distribution of the emitting electrons, resulting from the action of multiple shocks \citep{zech21} or acceleration by turbulence \citep{tavecchio22,sciaccaluga22}. At any rate, it is clear that the phenomenology of EHBL gives us access to one of the most extreme facets of acceleration in jets. {\it IXPE} observed the prototypical source of this class, 1ES0229+200, revealing the most energy-dependent degree of polarization among HSP, with 18\% in X-rays and 3\% in the optical band (as measured by NOT and OSN telescopes) \citep{ehlert2023}. The pronounced chromaticity displayed by the polarization could possibly be related to the extreme nature of these sources. 

In order to enlarge the sample of EHBL with polarimetric information and to investigate the possible connections with their peculiar phenomenology, we asked for {\it IXPE} pointings of two X-ray bright EHBL, namely 1ES 1101-232 and RGB J0710+591 \citep{costamante18}. To ensure a good coverage of the synchrotron peak and its polarimetric properties we complemented the {\it IXPE} observations with contemporaneous mm, optical, and X-ray observations with {\it Swift} and {\it NuStar}. In this paper we report the multiwavelength data, including polarimetric measurements, for the two sources and we discuss the results in the framework of the stratified shock scenario.  

The paper is organized as follows: in Sect. 2 we present the observations and the data analysis, in Sect. 3 we present the results and we discuss our findings in Sect. 4.

Throughout the paper, the following cosmological parameters are assumed: $H_0=67{\rm\;km\;s}^{-1}{\rm\; Mpc}^{-1}$, $\Omega_{\rm M}=0.3$, $\Omega_{\Lambda}=0.7$ (e.g. \citealt{Planck20}).

\section{Observations and data analysis}
\label{sec:data}

\subsection{{\it IXPE}}

\ixpe observations of two EHBLs were conducted on 1ES~1101-232 (Obs. ID 03006901, from 2024-11-28 to 2024-12-02 for $\sim$190 ks) and RGB~J0710+591 (Obs. ID 03007099, from 2024-03-20 to 2024-04-14 for $\sim$285 ks). X-ray polarization for each observation was estimated using both the event-by-event Stokes parameter analysis \citep[][the so-called model-independent analysis]{2015APh....68...45K} and the spectropolarimetric analysis \citep{2017ApJ...838...72S}. For the event-by-event Stokes parameter analysis, the polarization degree (\pdx) and polarization angle (\pax) were derived from the normalized Stokes parameters $q$ and $u$, obtained from the \texttt{pcube} products generated by the \texttt{xpbin} task in the \ixpeobssim software \citep{2022SoftX..1901194B}, using the relations {\small $\Pi_X = \sqrt{q^2 + u^2}$} and {\small $\psi_X = \frac{1}{2} \tan^{-1}(u/q)$}. The spectropolarimetric analysis was performed using the conventional X-ray spectral fitting tool \xspec \citep[version 12;][]{1996ASPC..101...17A} on the \I, \Q, and \U spectra. We modeled the spectra using \logpar model \cite{2004A&A...413..489M} to reproduce the synchrotron emission from the jets, along with a constant polarization model (\polconst) to estimate the polarization properties. The \texttt{pivotE} parameter was fixed at 2 keV.  Additionally, we accounted for cross-calibration factors among the three detector units (DUs) using \const, and modeled Galactic absorption with \tbab, adopting the weighted average column density values from \citet{HI4PI2016}, with $N_H = 5.13 \times 10^{20} \rm{cm}^{-2}$ for 1ES~1101-232 and $4.35 \times 10^{20} \rm{cm}^{-2}$ for RGB~J0710+591. The \texttt{willm} model was used to account for metal abundances \citep{2000ApJ...542..914W}. Error estimation in both analysis methods followed the guidelines provided by the \ixpe Science Team\footnote{\url{https://ixpe.msfc.nasa.gov/for_scientists/documentation/IXPE_Stats-Advice.pdf}}. The background rejection algorithm \citep{2022AJ....163..170D} was also applied to improve the sensitivity of the measurements. The overall data processing followed the procedures described in \citet{2024A&A...681A..12K}, which were applied in previous \ixpe observations of blazars. Regarding calibration, we employed the most recent versions of the \ixpe XRT calibration database (version 20241028) and the \ixpe GPD calibration database (version 20250225). All processing was carried out using \ixpeobssim version 31.0.3 and HEASoft version 6.34. For each source, both the event-by-event Stokes parameter analysis and the spectropolarimetric analysis were independently cross-validated, yielding results that are statistically consistent within the uncertainties. \citep[e.g.,][]{2024A&A...681A..12K}

\begin{figure*}[ht!]
\centering
\includegraphics[width=0.48\linewidth]{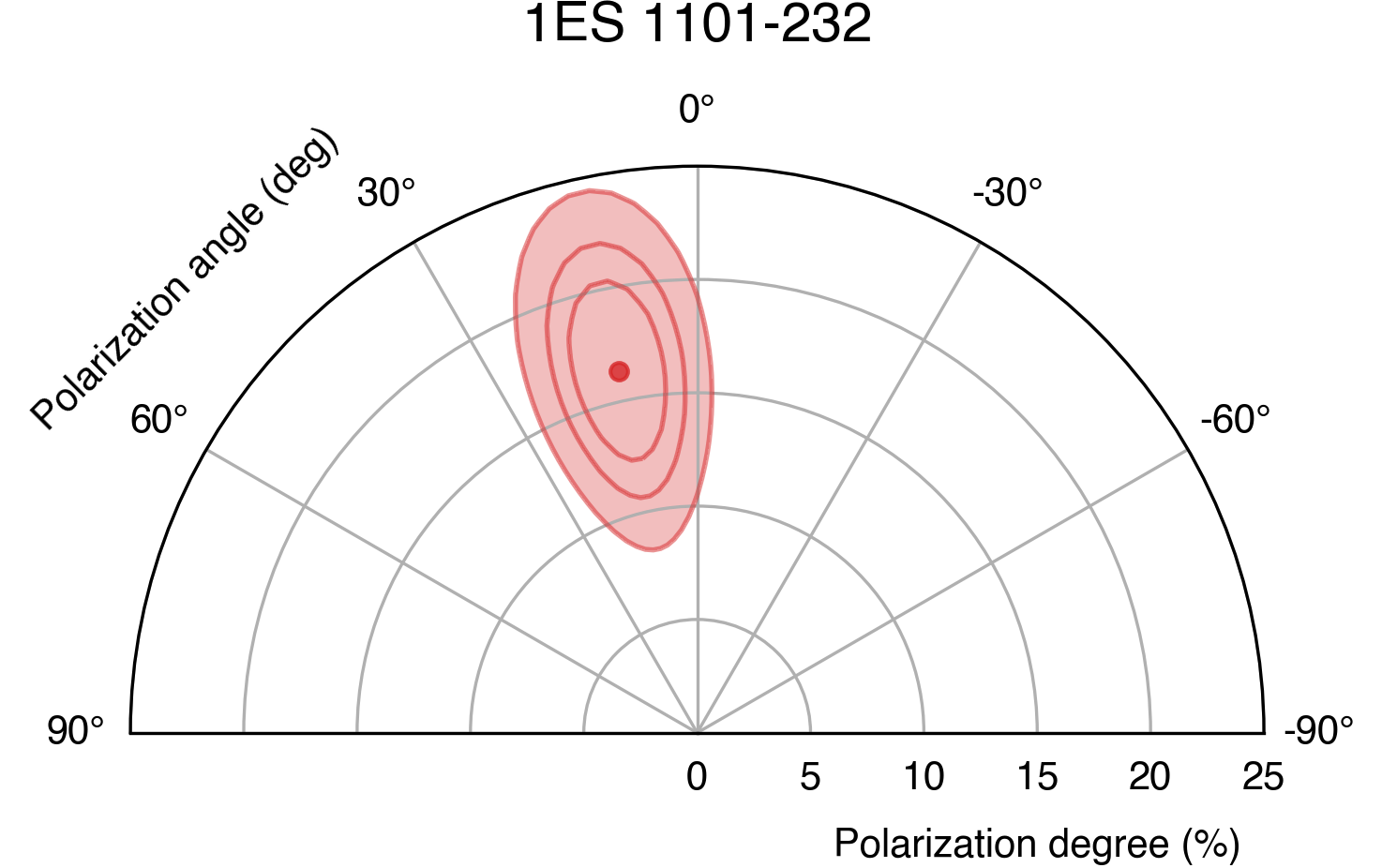}
\includegraphics[width=0.48\linewidth]{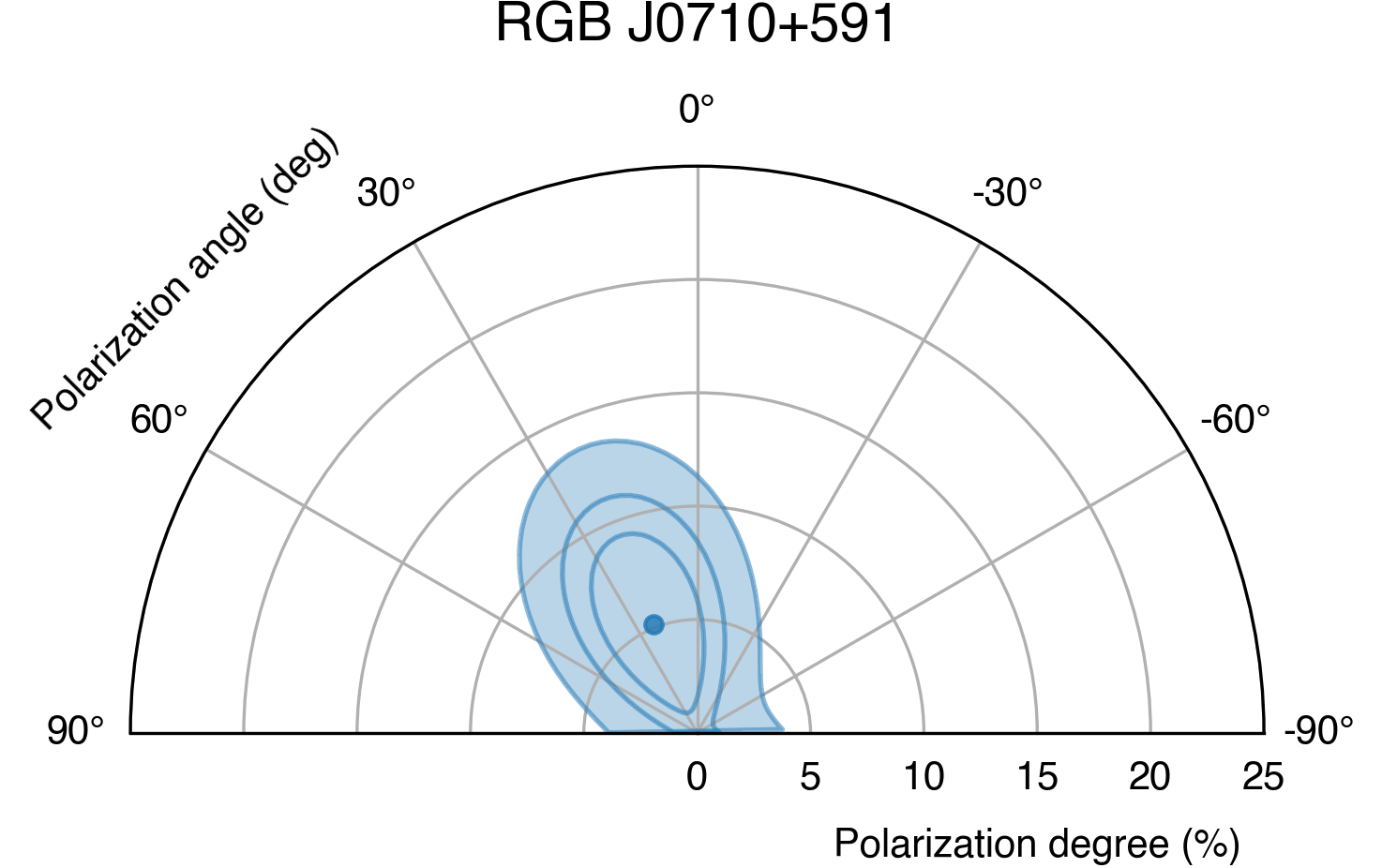}
\caption{Detection significance of X-ray polarization for 1ES~1101-232 (left) and RGB~J0710+591 (right), measured from the spectropolarimetric analysis. The tangential direction indicates the polarization angle, while the radial direction corresponds to the polarization degree. The colored dot at the center of each contour denotes the measured X-ray polarization. Each contour plot represents the detection confidence levels at 68\%, 90\%, and 99\%.}
\label{fig:polcont}
\end{figure*}

First of all, the time-averaged polarization was measured over the full duration of each observation in the 2–8 keV energy band. For 1ES~1101-232, the \pcube analysis yielded a polarization degree of \pdx = 15.9\% $\pm$ 3.8\% and a polarization angle of \pax = 15.7\degr $\pm$ 6.8\degr. The spectropolarimetric analysis constrained the polarization to \pdx = 16.3\% $\pm$ 2.7\% and \pax = 12.2\degr $\pm$ 4.7\degr. For RGB~J0710+591, the X-ray polarization was not significantly detected. The \pcube analysis provided a 99\% confidence level upper limit of \pdx $<$ 13.0\%, while the spectropolarimetric analysis yielded a slightly tighter, but consistent, 99\% confidence level upper limit of \pdx $<$ 11.6\%. Figure~\ref{fig:polcont} illustrates the measured X-ray polarization from the spectropolarimetric analysis, showing the detection significance in the polarization degree and angle parameter space for each observation. The spectral fitting results for each observation are summarized in Table~\ref{tab:xspec}.

\begin{table}[t]
\centering
\scriptsize
\caption{Best-fit parameters from the spectropolarimetric analyses of 1ES~1101-232 and RGB~J0710+591.}
\label{tab:xspec}
\resizebox{\columnwidth}{!}{
\begin{tabular}{llcc}
\hline\hline
\noalign{\smallskip}
& & 1ES~1101-232 & RGB~J0710+591 \\
\hline
\noalign{\smallskip}
Component & Parameter & value ($\pm 1 \sigma$) & value ($\pm 1 \sigma$) \\
\noalign{\smallskip}
\hline
\noalign{\smallskip}
\texttt{logpar} & alpha & 2.26 $\pm$ 0.08 & 2.17 $\pm$ 0.08 \\
 & beta & 0.54 $\pm$ 0.19 & 0.32 $\pm$ 0.19 \\
 & pivotE & 2.0  (\textcolor{magenta}{f}) & 2.0  (\textcolor{magenta}{f})\\
 & norm & 0.003 $\pm$ 4.e.-5 & 0.002 $\pm$ 3.e-5 \\
\noalign{\smallskip}
\hline
\noalign{\smallskip}
\texttt{polconst} & \pdx ($\%$) & 16.4 $\pm$ 2.7 & $<$ 13.5 \\
 & \pax (\degr) & 12.0 $\pm$ 4.7 & unconstrained  \\
\noalign{\smallskip}
\hline
\noalign{\smallskip}
\texttt{tbabs} & $N_H$ ($10^{22} {\rm cm^{-2}}$) & 5.13e-2 (\textcolor{magenta}{f}) & 4.35e-2 (\textcolor{magenta}{f}) \\
\noalign{\smallskip}
\hline
\noalign{\smallskip}
\texttt{constant} & IXPE DU 1 & 1 (\textcolor{magenta}{f}) & 1 (\textcolor{magenta}{f}) \\
 & IXPE DU 2 & 0.99 $\pm$ 0.01 & 1.03 $\pm$ 0.02 \\
 & IXPE DU 3 & 0.99 $\pm$ 0.01 & 1.00 $\pm$ 0.02 \\
\noalign{\smallskip}
\hline
\noalign{\smallskip}
\multicolumn{2}{l}{$\chi^2$ / ${\rm d.o.f.}$} & 356.01/389 & 335.3/393 \\
\noalign{\smallskip}
\hline
\noalign{\smallskip}
\multicolumn{2}{l}{Flux$_{2-8\,\rm{keV}}$ ($10^{-11}$\ergsc)} & 1.93$\pm$0.03 & 1.22$\pm$0.02 \\
\noalign{\smallskip}
\hline
\end{tabular}
}
\tablefoot{(\textcolor{magenta}{f}) denotes fixed parameters.}
\end{table}

\begin{figure}[ht!]
\centering
\includegraphics[width=1.\linewidth]{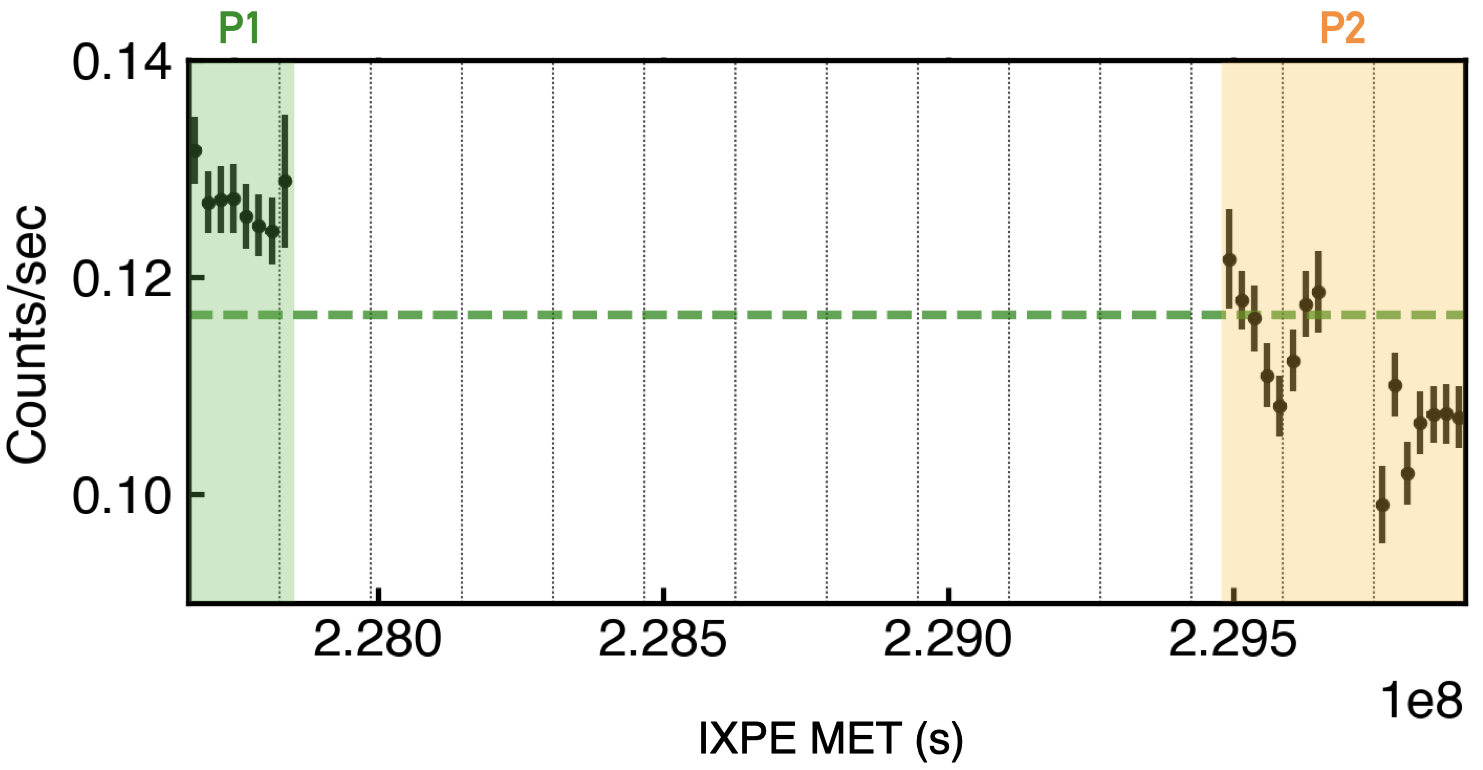}
\caption{Light curve of the \ixpe observation of RGB~J0710+591. The green and orange shaded areas indicate periods 1 and 2, respectively, which were split based on the observation gap in the middle. The central green dashed line represents the average count rate of RGB~J0710+591 during the \ixpe observation.}
\label{fig:lc_rgb}
\end{figure}

Furthermore, in order to investigate the potential variability of polarization over time, we conducted a time-resolved analysis. In this test, we estimated the null hypothesis probability for fitting a constant model to the $q$ and $u$ Stokes parameters independently, dividing the entire observation according to integer ratios from 2 to 15, following the method presented in \citet{2024A&A...681A..12K}. For both observations, the null hypothesis probabilities for the constant model fits to the $q$ and $u$ Stokes parameters were greater than 1\%, indicating that no significant deviation from a constant polarization signal was found. Thus, the observed polarization variations are consistent with random fluctuations. In particular, as an additional test for RGB~J0710+591, considering the observational gap in the middle of the exposure as shown in Figure \ref{fig:lc_rgb}, we attempted to split the dataset into period 1 (P1; green shaded area) and 2 (P2; orange shaded area), and measure the time-averaged polarization separately. However, despite a slight increase in significance in P2, the polarization remained unconstrained.

We also performed an energy-resolved analysis by subdividing the \ixpe energy range (2--8 keV) into narrower intervals, beginning with two bins (2--4 keV and 4--8 keV), then three bins (2--4, 4--6, and 6--8 keV), and extending up to 12 bins in total. The results showed no statistically significant variation relative to a constant model, indicating that the X-ray polarization remains consistent across energy and shows no energy dependence. Notably, in the 3--4 keV range, we found that the \pcube analysis of RGB~J0710+591 detected a polarization signal corresponding to \pdx=16.6\%$\pm$5.2\% and \pax=32.6\degr$\pm$9.0\degr at the 99.4\% confidence level. This signal, however, was not reproduced in the spectropolarimetric analysis using \xspec, likely because \xspec estimates the uncertainties based on chi-square statistics, that provides a low significance when using such a narrow energy bin (3--4 keV) with large uncertainty.

\begin{figure}[ht!]
\centering
\includegraphics[width=1.\linewidth]{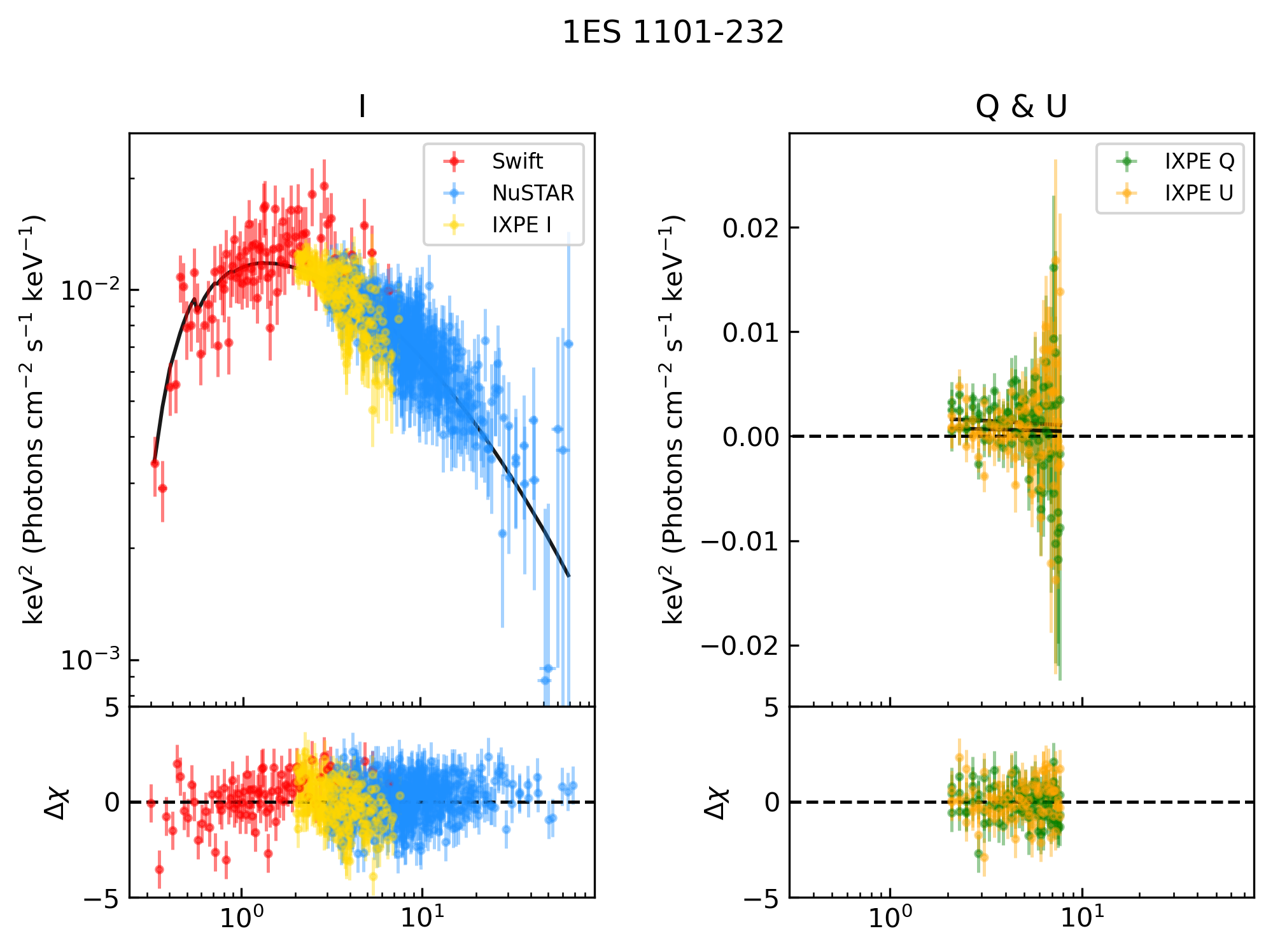}
\caption{A joint spectropolarimetric analysis of 1ES~1101-232. The left panel presents the energy flux spectra, expressed as photon flux multiplied by energy squared, for the Swift (red), IXPE I (yellow), and NuSTAR (blue) data together with the corresponding data-model residuals. The black line indicates the best-fit model. The right panel displays the IXPE Q and U spectra, shown in green and orange, respectively, together with their deviations from the best fit model.}
\label{fig:xspec_joint_1es}
\end{figure}

\begin{figure}[ht!]
\centering
\includegraphics[width=1.\linewidth]{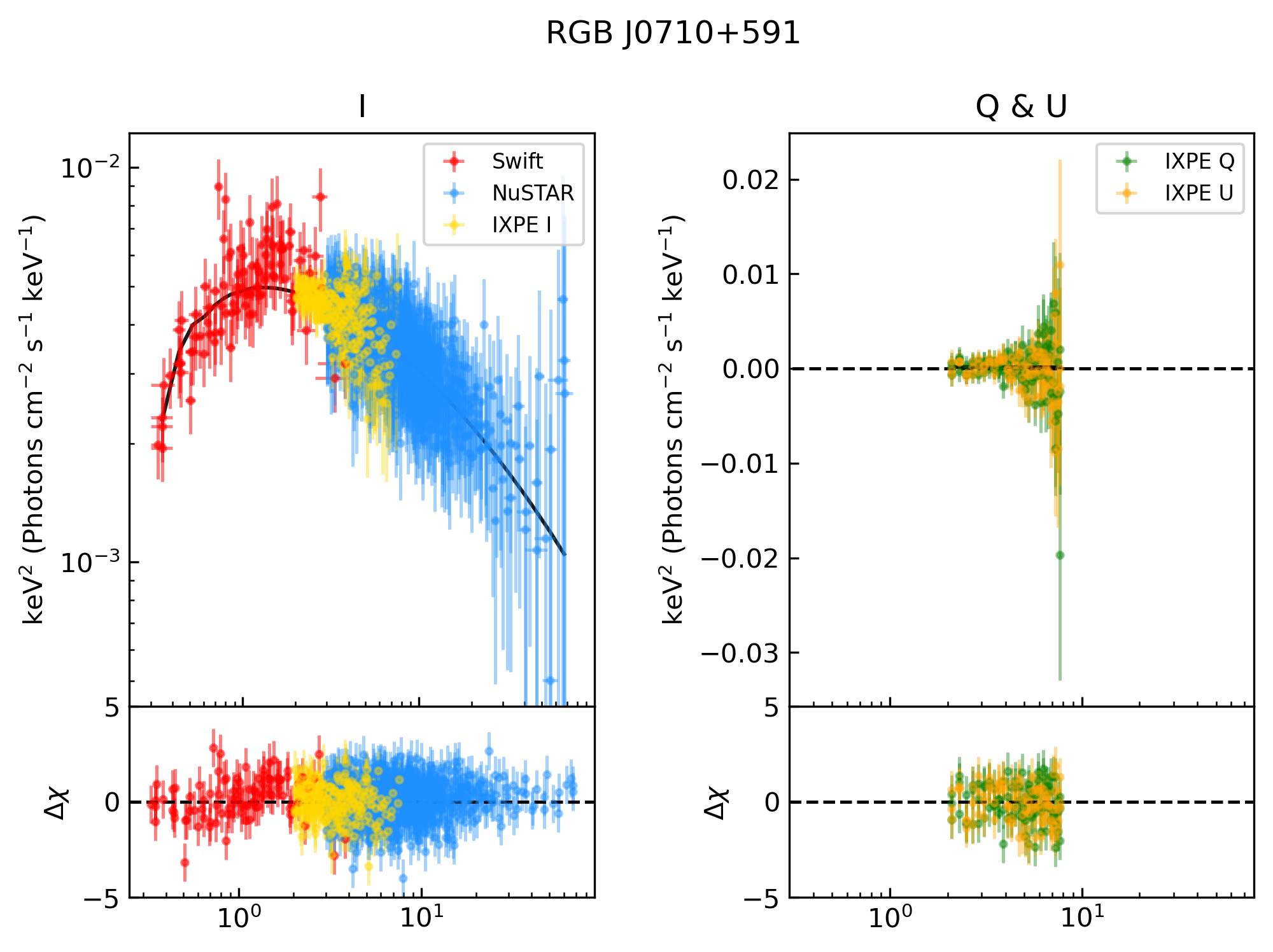}
\caption{A joint spectropolarimetric analysis of RGB~J0710+591. The panels follow the same format as Figure~\ref{fig:xspec_joint_1es}.}
\label{fig:xspec_joint_rgb}
\end{figure}

We performed a joint spectropolarimetric analysis adding {\it Swift} and {\it NuSTAR} (for the respective analyses and details see below). The same spectral model used in the IXPE-only analysis has been applied, \texttt{constant$\times$tbabs$\times$polconst$\times$logpar}. We fixed the \const as unity for {\it Swift} spectra as the reference spectrum and allowed it to vary for the others. Table~\ref{tab:xspec_joint} presents the results of the joint spectral model and Figs. \ref{fig:xspec_joint_1es} and \ref{fig:xspec_joint_rgb} show the spectral fits. As a result, we obtained slightly improved constraints on both spectral and polarization properties. While the alpha parameter in \logpar, which indicates the spectral slope, was deviated in the case of 1ES~1101-232, other parameters, including polarization properties, were consistent within uncertainties. The polarization degree and angle obtained from the joint fitting analysis were \pdx=15.7\%$\pm$2.6\% and \pax=12.2\degr$\pm$4.7\degr for 1ES~1101-232, and the 99\% confidence level upper limit of \pdx $<$ 12.2\% for RGB~J0710+591.

\begin{table}[t]
\centering
\scriptsize
\caption{Best-fit parameters from the joint spectropolarimetric analyses of 1ES~1101-232 and RGB~J0710+591. \label{tab:xspec_joint}}
\label{tab:xspec}
\resizebox{\columnwidth}{!}{
\begin{tabular}{llcc}
\hline\hline
\noalign{\smallskip}
& & 1ES~1101-232 & RGB~J0710+591 \\
\hline
\noalign{\smallskip}
Component & Parameter & value ($\pm 1 \sigma$) & value ($\pm 1 \sigma$) \\
\noalign{\smallskip}
\hline
\noalign{\smallskip}
\texttt{logpar} & alpha & 2.19 $\pm$ 0.02 & 2.16 $\pm$ 0.02 \\
 & beta & 0.23 $\pm$ 0.02 & 0.18 $\pm$ 0.02 \\
 & pivotE & 2.0  (\textcolor{magenta}{f}) & 2.0  (\textcolor{magenta}{f})\\
 & norm & 0.003 $\pm$ 8.e.-5 & 0.001 $\pm$ 3.e-5 \\
\noalign{\smallskip}
\hline
\noalign{\smallskip}
\texttt{polconst} & \pdx ($\%$) & 15.7 $\pm$ 2.6 & $<$ 12.1 \\
 & \pax (\degr) & 12.2 $\pm$ 4.7 & unconstrained  \\
\noalign{\smallskip}
\hline
\noalign{\smallskip}
\texttt{tbabs} & $N_H$ ($10^{22} {\rm cm^{-2}}$) & 5.13e-2 (\textcolor{magenta}{f}) & 4.35e-2 (\textcolor{magenta}{f}) \\
\noalign{\smallskip}
\hline
\noalign{\smallskip}
\texttt{constant} & Swift & 1 (\textcolor{magenta}{f}) & 1 (\textcolor{magenta}{f}) \\
 & NuSTAR FPMA & 1.16 $\pm$ 0.04 & 1.65 $\pm$ 0.06 \\
 & NuSTAR FPMB & 1.19 $\pm$ 0.04 & 1.69 $\pm$ 0.06 \\
 & IXPE DU 1 & 0.98 $\pm$ 0.03 & 1.36 $\pm$ 0.04 \\
 & IXPE DU 2 & 0.96 $\pm$ 0.03 & 1.40 $\pm$ 0.04 \\
 & IXPE DU 3 & 0.96 $\pm$ 0.03 & 1.35 $\pm$ 0.04 \\
\noalign{\smallskip}
\hline
\noalign{\smallskip}
\multicolumn{2}{l}{$\chi^2$ / ${\rm d.o.f.}$} & 1028.72/954 & 1210.31/1340 \\
\noalign{\smallskip}
\hline
\noalign{\smallskip}
\multicolumn{2}{l}{Flux$_{2-8\,\rm{keV}}$ ($10^{-11}$\ergsc)} & 2.12$\pm$0.04 & 1.01$\pm$0.04 \\
\noalign{\smallskip}
\hline
\end{tabular}
}
\tablefoot{(\textcolor{magenta}{f}) denotes fixed parameters.}
\end{table}

\subsection{{\it Swift}/XRT}

The {\it Swift} satellite observed the two sources simultaneously to the \ixpe observations: 1ES1101-232 on 2024-11-28,29 (ObsIDs: 00035013052-00035013053) and 2024-12-01,02,04 (ObsIDa: 00035013055-00035013057), and RGBJ0710+591 on 2024-04-04,09,11,13,15 (ObsID: 00031356080-00031356082, 00031356084, 00031356085). 
Data of the X-ray Telescope \citep[XRT;][]{burrows05} were downloaded from HEASARC public archive, processed
with the Swift XRT Data Analysis Software (SWXRTDAS; version 3.7.0) and the relevant software included in the package HEASOFT v.\ 6.36.
The calibration database was updated on 2025 June 09. 

The total exposure of 1ES1101-232 on the XRT was of 4.7 ks. 
The fit was performed with a simple power-law model and Galactic absorption \citep[$N_H=7.03\times10^{20} \rm cm^{-2}$;][]{willingale13}, and using unbinned likelihood \citep{cash79}.
The output parameters of the model were $\Gamma = 2.01\pm0.05$ and an integrated observed flux $F_{\rm0.3-10 keV} = 6.21^{+0.17}_{-0.16} \times 10^{-11}\rm  erg\, cm^{-2} s^{-1}$. 
The individual observations were also fitted separately, resulting in significantly consistent flux and photon index values.

RGBJ0710+591 was observed by XRT for a total of 9.8 ks.
The data were fitted under the same assumptions as 1ES1101-232, with Galactic absorption $N_H=5.15\times10^{20} \rm cm^{-2}$ and a single power-law.
No variability is suggested by fitting the individual observations, and therefore we preferred to combine all of them in a single fit. 
The data are best fitted with $\Gamma = 1.86\pm0.04$ and  $F_{\rm0.3-10 keV} = 3.35^{+0.07}_{-0.05} \times 10^{-11}\rm  erg\, cm^{-2} s^{-1}$.

\subsection{{\it NuSTAR}}

The simultaneous {\it NuSTAR} observations were undertaken during the \ixpe{} observations for both targets (1ES1101-232: Obs. ID 61001017002; RGBJ0710+591: Obs. IDs 91001607002 and 91001607004) in order to better constrain the spectral properties by covering an extended energy range together with \ixpe{}. {\it NuSTAR} data reduction was performed using \texttt{nustardas} version 2.1.5 and CALDB version 20241216 following the procedures described in the {\it NuSTAR} data analysis software users guide\footnote{\href{https://heasarc.gsfc.nasa.gov/docs/nustar/analysis/nustar_swguide.pdf}{https://heasarc.gsfc.nasa.gov/docs/nustar/analysis/nustar\_swguide.pdf}}. For both the FPMA and FPMB detectors, the level 2 event files were generated using the \texttt{nupipeline} task, while high-level scientific data products, including spectral extraction were produced using the \texttt{nuproducts} tool. A circular source region with a $1$\arcmin{} radius and a circular background region with a $2$\arcmin{} radius were applied for the source and background extraction. Each spectrum was grouped to have a minimum of 30 counts per bin. In order to enable simultaneous spectropolarimetric analysis with \ixpe, ``XFLT0001 Stokes:0'' keyword was added to the header, allowing \xspec{} to identify the spectra as Stokes $I$ spectra.

\subsection{Optical, millimeter-wave and radio observations}
\begin{figure}
    \centering
    \includegraphics[width=\linewidth]{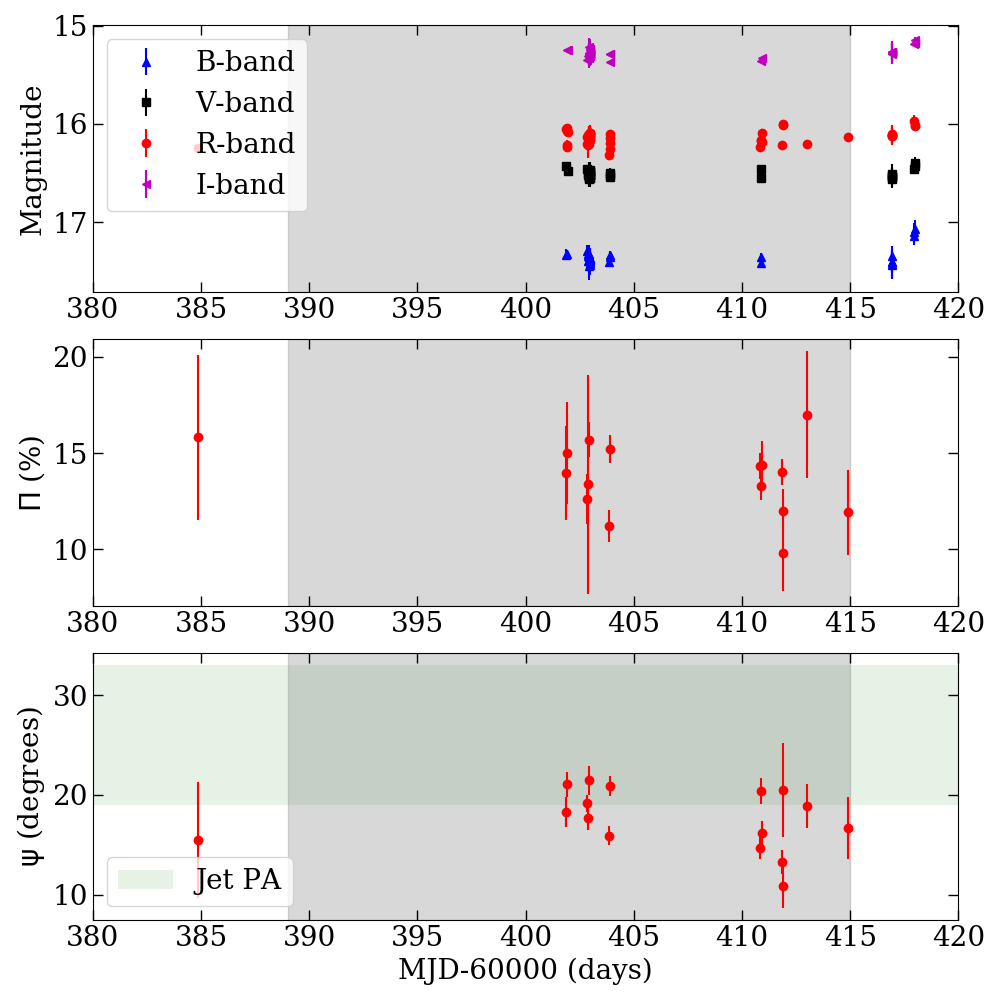}
    \caption{Optical polarization observations of RGB J0710+591. The top panel shows the brightness in different optical bands, the middle panel the polarization degree in the R-band, and the bottom panel the polarization angle. The grey shaded area marks the duration of the IXPE observation, whereas the horizontal green band marks the direction of jet as projected on the sky.}
    \label{fig:MWL_0710}
\end{figure}

\begin{figure}
    \centering
    \includegraphics[width=\linewidth]{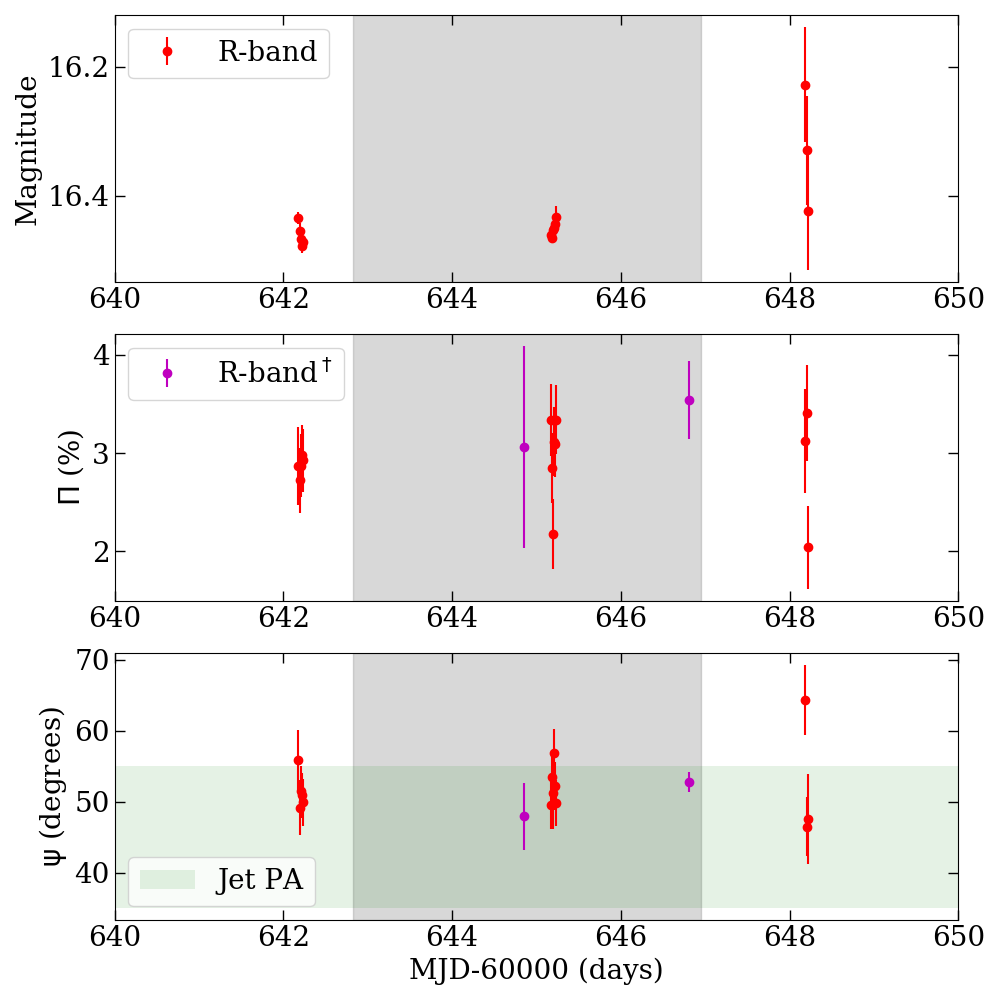}
    \caption{Optical polarization observations of 1ES 1101-232. Panels and symbols as in Fig. \ref{fig:MWL_0710}. The R-band$^\dagger$ measurements {\bf (purple symbols)} have not been corrected for the host galaxy contribution.}
    \label{fig:MWL_1101}
\end{figure}

Optical and radio/mm polarization observations were performed for both sources using the Calar Alto Faint Object Spectrograph (CAFOS, \citealp{escudero2024iop4}) on the Calar Alto 2.2m Telescope, the DIPOL-1 polarimeter at the Sierra Nevada Observatory \citep[DIPOL-1,][]{otero2024} using the IOP4 data reduction pipeline,  \citep{juan_escudero:2023,escudero2024iop4}, the Alhambra Faint Object Spectrograph and Camera (ALFOSC) at the Nordic Optical Telescope (NOT, \citealp{Nilsson2018}), the KANATA telescope using the Hiroshima Optical and Near-InfraRed camera (HONIR, \citealp{kawabata_new_1999,akitaya_honir_2014}), the 60~cm telescope at the Belogradchik Observatory \citep{Bachev2024}, the SMA Monitoring of AGN with POLarization (SMAPOL) program \citep{Myserlis2025},  and finally the Effelsberg 100-m telescope through the monitoring the Stokes Q, U, I, and V emission of AGN jets in Radio  (QUIVER, \citealp{kraus2003,myserlis2018,Myserlis2025}) and the Tev Effelsberg Long-term Agn MONitoring (TELAMON, \citealp{Eppel2024}) programs. All the optical observations were reduced following standard calibration procedures using polarized and unpolarized standard stars to account for instrumental polarization. Where possible, the R-band measurements have been corrected for the depolarizing effect of the host-galaxy flux using the galaxy profiles from \cite{Falomo2000,Nilsson2007} following \cite{Hovatta2016}. The optical polarization observations are shown in Fig. \ref{fig:MWL_0710} and \ref{fig:MWL_1101}.
Note that, although some of the measurements relative to 1ES 1101-232 have not been corrected, the effect on the net polarization is small.

RGB J0710+591 shows a very high degree of optical polarization with a median and standard deviation over the IXPE observation of $\Pi_O=13.9\%$ and $\sigma_{\Pi_O}=1.8\%$. On the contrary, the radio polarization observations yielded only upper limits (at the 3$\sigma$ level) of $<15\%$ at 2.5~GHz,  $<9.8\%$ at 4.8~GHz, $<3\%$ at 6.6~GHz, $<30\%$ at 10.4~GHz, $<3\%$ at 13.6~GHz, and $<6.9\%$ at 225.5~GHz. The median optical polarization angle for the duration of the IXPE is at $\rm\psi_O=18.3$ degrees with a standard deviation of $\rm\sigma_{\psi_O}=3$. Estimates of the direction of the jet vary from 10 degrees to 50 degrees \cite{Wu2012,Hovatta2016, Plavin2022} with an average of 26 degrees and a wide opening angle of 45 degrees \cite{Wu2012}. Although the VLBI observations are not contemporaneous with the optical observations the latter seem to be fairly aligned with the jet axis, as is often the case for HSP sources \citep{liodakis22,Kouch2024,Capecchiacci2025}.

1ES 1101-232 shows much lower polarization than  RGB J0710+591 with a median and standard deviation over the IXPE observation of $\Pi_O=3.1\%$ and $\sigma_{\Pi_O}=0.4\%$ and a median polarization angle of $\rm\psi_O=51.7$ degrees with a standard deviation of $\rm\sigma_{\psi_O}=3.4$. Also in this case, the optical  $\rm\psi_O$ is fairly aligned with the jet axis \citep{Benke2024}.

\section{Results}
\label{sec:results}

As expected because of the small variability characterizing EHBL, the spectral parameters of the fitted X-ray spectrum (including for the two sources {\it Swift} and {\it NuSTAR}) are very similar to those reported by \cite{costamante18}, which also included {\it NuSTAR} data. In both cases the synchrotron component, well reproduced by a log-parabolic shape, peaks around 1 keV. Therefore, IXPE measurements probe the high-energy tail of the synchrotron emission just after the peak (see Figs.\ref{fig:xspec_joint_1es}-\ref{fig:xspec_joint_rgb}).

For RGB J0710+591 we derive a relatively stringent upper limit of the polarization in the X-ray band, even splitting the data in energy and time. Quite interestingly, the upper limit is below (or at least comparable, considering a confidence level of 3$\sigma$) the (galaxy-corrected) optical degree of polarization. This measurements are in contrast with all previous results obtained for BL Lacs of the HBL and EHBL type, which are characterized by a pronounced increase of the degree of polarization with frequency, with typically $\Pi_X\gtrsim 2\, \Pi_O$ \citep[e.g.][]{Kim24,Marscher24}. The only exception is the HBL 1ES 1959+650, that during a pointing in June 2022 displayed $\Pi_O=4.7$ and $\Pi_X<5.1$ \citep{Errando2024}\footnote{However, when the IXPE observation is split into four time bins, the third one gives a significative detection with $\Pi_X=7.5$ \citep{Errando2024}.}. Note that the prototypical EHBL 1ES 0229+220 represents one of the most extreme cases of chromaticity, with $\Pi_X/\Pi_O\sim 6$ \citep{ehlert2023}.

1ES 1101-232, on the other hand, appears in line with typical BL Lacs, with a highly polarized X-ray emission, $\Pi_X\simeq16\%$ and a rather modest polarization in the optical band, $\Pi_O\simeq 3\%$, which translate in a ratio $\Pi_X/\Pi_O\sim 5$. 

In both sources the optical EVPA is consistent with the jet PA, again in agreement with the typical behavior of HBL \citep[e.g.][]{Kim24,Marscher24}. For 1ES 1101-232, $\Psi_X$ instead lies in the range 8-16 degrees, inconsistent with the optical EVPA, $\Psi_O\sim 51$ degrees, and with the jet PA. Interestingly, while most of the BL Lac show a substantial agreement between $\Psi_X$ and the jet PA \cite[e.g., ][]{Capecchiacci2025}, also in the prototypical EHBL 1ES 0229+220 the X-ray EVPA is misaligned with respect to the jet PA by about 50 degrees. In considering these results it is important to note that the jet could bend by tens of degrees between the inner region (where the X-ray and optical emission are likely produced) and the outer region imaged in the radio band \citep[e.g.][]{digesu22}. Moreover, jets can change PA with time \citep{Kostrichkin25}, while in most of the cases the radio and optical and X-rays observations are not simultaneous.

\section{Discussion}
\label{sec:discussion}

Our polarimetric measurements, referring to two EHBL that share quite similar spectral energy distributions \citep{costamante18}, add complexity to the observational framework of extreme blazars. 

1ES 1101-232, which displays high polarization in the X-rays accompanied by a much more modest degree of polarization in the optical, follows the trend commonly observed in these types of sources. The ratio of X-ray to optical polarization, $\Pi_X/\Pi_O\simeq 5.2$, is very similar to the EHBL prototype 1ES 0229+200, with $\Pi_X/\Pi_O\simeq 5.6$ \citep{ehlert2023}\footnote{Presently, the most extreme chromaticity has been displayed by PKS 2155-304, which reached an extreme $\Pi_X/\Pi_O\simeq 7.2$ \citep{Kouch2024}.}.
On the other hand, the upper limit obtained for RGB J0710+591, together with the large optical polarization (with some datapoints reaching 15\%, the largest value recorded among HBL and EHBL monitored by {\rm IXPE}, \citealt{Kim24}) clearly demonstrates that not all sources follow the simple increasing trend of the degree of polarization with frequency.

\subsection{A stratified shock model}

The observed strong chromaticity of the degree of polarization is often proposed as a support for the stratified shock scenario \citep[e.g.][]{angelakis16,tavecchio18,tavecchio21,marscher22}. In this model, particles are thought to be injected at a shock in the jet and further advected by the downstream flow. Owned to the energy-dependent cooling time of the electrons, X-rays are produced only very close to the shock, where the field component orthogonal to the shock could be more coherent due to compression and kinetic effects. Electrons producing the optical emission, on the other hand, are characterized by a larger cooling length and therefore fill a large portion of the post-shock fluid, likely experiencing more turbulent and less coherent fields, therefore resulting in a small degree of polarization.

\begin{figure}
\hspace{-0.2truecm}
\includegraphics[width=1.06\linewidth]{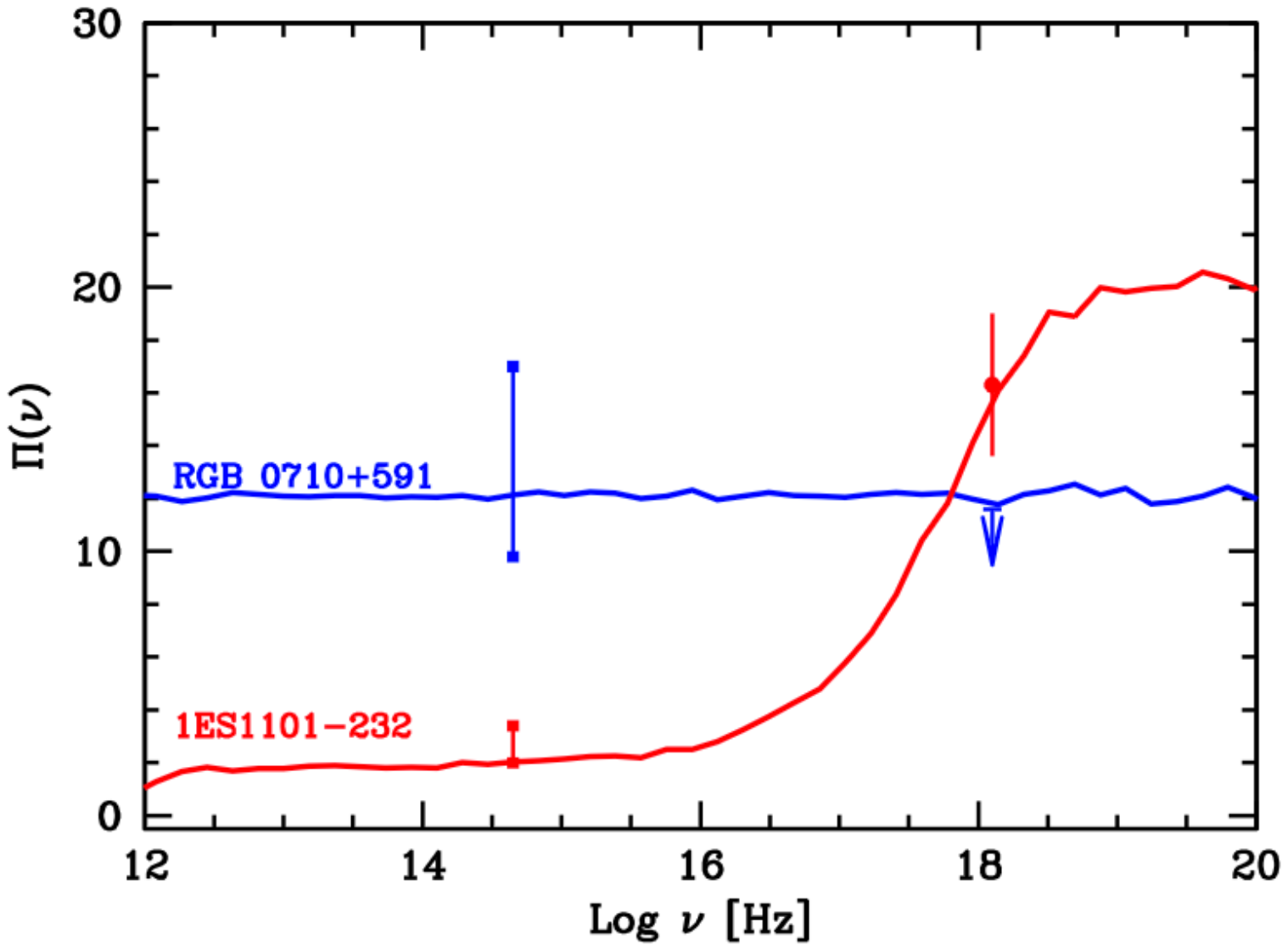}
\caption{Optical and X-ray degree of polarization measured for the two sources. For the optical band, the bar reports the range of the measured values. The curves show two realization of the stratified shock scenario described in the text.}
\label{fig:model}
\end{figure}

Despite its simplicity, the stratified shock scenario is sufficiently versatile and it can reproduce the different behavior of the two sources. In Fig. \ref{fig:model} we report the degree of polarization measured in the optical and in the X-ray band for both sources (for the optical band, where we have several measurements, we report the range of the recorded values), together with two realizations of an improved version of the \cite{tavecchio18} stratified shock model (see Appendix A for details and parameters). For 1ES 1101-232 the model reproduces the standard chromatic behavior. For RGB J0710+591 it is instead possible to reproduce the observed optical data and the X-ray upper limit considering a jet closely oriented toward the observer and a shallower profile for the orthogonal magnetic field. In this case the $\Pi(\nu)$ relation shows no chromaticity (see Appendix A). It is important to remark that the model can reproduce the data because the X-ray upper limit is almost at the same level of the measured optical polarization. Lower upper limits (or detection) would be very challenging for this scenario.

Although the model shown above can satisfactorily reproduce the data, it is built on highly idealized and simplified hypotheses. By means of dedicated MHD simulations, \cite{Sciaccaluga25} showed that this simple model can be shaped in a realistic scenario considering the shocks produced in an underpressured jet recollimated by the external medium \citep{KomissarovFalle97}. The model is able to reproduce the commonly observed chromaticity of the polarization. However, the same study also reveals that the properties of the polarization, in particular its chromaticity, depend on the specific details of the system. In particular, the ratio of the poloidal and the toroidal components of the helical magnetic field of the jet (i.e. the pitch) has a great impact. Interestingly, when the toroidal field is dominating, one obtains a reversed chromaticity, i.e. the degree of polarization decreases with the frequency. This behavior is mainly dictated by the asymmetries related to the complex pattern of the Doppler boosting of the radiation produced in the complex flow downstream of the recollimation shock. The asymmetries are much more pronounced at low frequencies \citep{Sciaccaluga25}, since the emitting electrons extend over a larger region. The X-ray emission, on the contrary, is produced in a compact region just after the shock, for which the boosting is much more uniform. Following these results, one could speculate that the reversed chromaticity of RGB J0710+591 is related to a toroidal field larger than typically found in the jet of the other blazars. 

\subsection{Other scenarios}

It is also interesting to contrast the case of RGB J0710+591 with the alternative model proposed by \cite{Bolis24}. In this model the chromaticity is not related to the energy stratification of the emission region, but it is naturally related to the geometry of the fields. Assuming a structure of the field self-consistently calculated in the theory of force-free, Poynting dominated jet, \cite{Bolis24} showed that one can obtain the observed chromaticity if electrons occupy the same region and the jet has a parabolic shape. The increase of the polarization with the frequency is a robust feature of this model and therefore the results on RGB J0710+591 are challenging to interpret without supplementary assumptions (for instance, multiple regions of emission).

An alternative approach, motivated by the standard model for jet production and acceleration \citep[e.g.][]{Komissarov2009} assumes that emitting particles are energized through magnetic reconnection, thought to be more efficient than shock acceleration when the plasma is highly magnetized \citep{Sironi15b}. Both MHD simulations of magnetic reconnection triggered by kink instability in highly magnetized jets \citep{bodo21} or PIC simulations of single current sheaths \citep{zhang20} display highly variable polarization (and flux) at both optical and X-rays energies, with a comparable averaged degree of polarization. However, a scenario based on these results seems incompatible with the EHBL phenomenology, in particular with the small level of variability characterizing these sources \citep{biteau20} and the small magnetization inferred through the modeling of the SED \citep{TavecchioGhisellini16}.

The results presented in this paper highlight the complexity of the polarimetric properties of the synchrotron emission from blazars and the necessity to have a good simultaneous multiwavelength coverage, extending from radio to X-rays. We also remark the importance of refined models able to catch the complexity of the phenomenology displayed by these sources, also including the polarimetric channel.

\begin{acknowledgements}
 We acknowledge financial support from a INAF Theory Grant 2024 (PI F.~Tavecchio). This work has been funded by the European Union-Next Generation EU, PRIN 2022 RFF M4C21.1 (2022C9TNNX). We acknowledge financial support from ASI grant I/004/11/6. We thank Dr. Brad Cenko and the Swift team for approving and carrying out the Swift ToO observations. Part of this work is based on archival data provided by the Space Science Data Center-ASI.The IAA-CSIC co-authors acknowledge financial support from the Spanish "Ministerio de Ciencia e Innovaci\'{o}n" (MCIN/AEI/ 10.13039/501100011033) through the Center of Excellence Severo Ochoa award for the Instituto de Astrof\'{i}sica de Andaluc\'{i}a-CSIC (CEX2021-001131-S), and through grants PID2019-107847RB-C44 and PID2022-139117NB-C44. Some of the data are based on observations collected at the Observatorio de Sierra Nevada; which is owned and operated by the Instituto de Astrof\'isica de Andaluc\'ia (IAA-CSIC), and at the Centro Astron\'{o}mico Hispano en Andaluc\'ia (CAHA); which is operated jointly by Junta de Andaluc\'{i}a and Consejo Superior de Investigaciones Cient\'{i}ficas (IAA-CSIC). The Submillimeter Array is a joint project between the Smithsonian Astrophysical Observatory and the Academia Sinica Institute of Astronomy and Astrophysics and is funded by the Smithsonian Institution and the Academia Sinica. Maunakea, the location of the SMA, is a culturally important site for the indigenous Hawaiian people; we are privileged to study the cosmos from its summit. E.L. was supported by Academy of Finland projects 317636 and 320045.  P.K. was supported by Academy of Finland projects 346071 and 345899. P.K. acknowledges support from the Mets\"ahovi Radio Observatory of Aalto University. This work was supported by JST, the establishment of university fellowships towards the creation of science technology innovation, Grant Number JPMJFS2129. This work was supported by Japan Society for the Promotion of Science (JSPS) KAKENHI Grant Numbers JP21H01137. This work was also partially supported by Optical and Near-Infrared Astronomy Inter-University Cooperation Program from the Ministry of Education, Culture, Sports, Science and Technology (MEXT) of Japan. We are grateful to the observation and operating members of Kanata Telescope. I.L was funded by the European Union ERC-2022-STG - BOOTES - 101076343. Views and opinions expressed are however those of the author(s) only and do not necessarily reflect those of the European Union or the European Research Council Executive Agency. Neither the European Union nor the granting authority can be held responsible for them. The data in this study include observations made with the Nordic Optical Telescope, owned in collaboration by the University of Turku and Aarhus University, and operated jointly by Aarhus University, the University of Turku and the University of Oslo, representing Denmark, Finland and Norway, the University of Iceland and Stockholm University at the Observatorio del Roque de los Muchachos, La Palma, Spain, of the Instituto de Astrofisica de Canarias. The data presented here were obtained in part with ALFOSC, which is provided by the Instituto de Astrof\'{\i}sica de Andaluc\'{\i}a (IAA) under a joint agreement with the University of Copenhagen and NOT. We acknowledge funding to support our NOT observations from the Finnish Centre for Astronomy with ESO (FINCA), University of Turku, Finland (Academy of Finland grant nr 306531). This research was partially supported by the Bulgarian National Science Fund of the Ministry of Education and Science under grants KP-06-H68/4 (2022), KP-06-H78/5 (2023) and KP-06-H88/4 (2024). Partly based on observations with the 100-m telescope of the MPIfR (Max-Planck-Institut f\"ur Radioastronomie) at Effelsberg. Observations with the 100-m radio telescope at Effelsberg have received funding from the European Union’s Horizon 2020 research and innovation programme under grant agreement No 101004719 (ORP). F.E., S.H., J.H., M.K., and F.R. acknowledge support from the Deutsche Forschungsgemeinschaft (DFG, grants 447572188, 434448349, 465409577). G. F. P. acknowledges support by the European Research Council advanced grant “M2FINDERS – Mapping Magnetic Fields with INterferometry Down to Event hoRizon Scales” (Grant No. 101018682). C.C. acknowledges support from the European Research Council (ERC) under the Horizon ERC Grants 2021 programme under grant agreement No. 101040021.
\end{acknowledgements}

\bibliographystyle{aa}
\bibliography{tavecchio}
\begin{appendix}

\section{An improved stratified shock model}

In this Appendix we briefly sketch the model that we have used to reproduce the multiwavelength polarimetric data of the two sources. 

\subsection{The model}

The model improves the original scenario described in \cite{tavecchio18}. In the following we highlight the main modifications.

The model assumes that relativistic electrons are injected at the front of a mildly relativistic shock normal to the jet axis, with downstream velocity (in the shock frame) $\beta_{\rm d, sh}\simeq1/3$ which propagates in a cold cylindrical jet with radius $r_j$.

The upstream jet flow is supposed to carry a weak field almost parallel to the jet axis with intensity $B_{\parallel}$. A magnetic field with lines forming a small angle with respect to the shock normal  (a configuration called a ``parallel" shock)  is prerequisite to have an efficient particle acceleration \citep[e.g.][]{Sironi15b}. Kinetic simulations show the formation of an intense field parallel to the shock, self-generated by streaming accelerating particles, slowly decaying with distance in the downstream flow. 
Following \cite{lemoine13}, for definiteness we assume that the intensity of the generated field decays along the jet flow following a phenomenological power law: 
\begin{equation}
B_{\perp}(d)= B_{\perp, 0} \left(1+\frac{d}{\lambda}\right)^{-m},
\label{eq:bperp}
\end{equation}
where $d$ is the distance in the downstream measured from the shock and the power law index $m$ is in the range $0.2-0.5$. With this parametrization, $\lambda$ plays the role of an effective decay length. Furthermore, we  account for the micro-turbulent  nature of the self-generated magnetic field by using a cell structure, each cell representing a coherence domain. At each point of the grid of $d$, we model $n_{\rm cells}=5\times 10^3$ equal cells assumed to uniformly fill the jet cross section (we checked that with a larger number of cells the results do not change). In each cell we specify the total magnetic field as the sum of the constant parallel field $B_{\parallel}=B_z$ and the orthogonal field $\mathbf{B_{\perp}}(d)$ (evaluated at distance of the cell from the shock, $d$), with components $B_x$, $B_y$, which we randomly select in each cell under the condition $B_x^2+B_y^2=B^2_{\perp}(d)$. The simulation reported in \cite{tavecchio18} suggests that $B_{\perp, 0}/B_{\parallel}\lesssim 10$.

Relativistic electrons accelerated at the shock are advected downstream by the flow, experiencing radiative losses due to the emission of synchrotron and IC radiation.
\cite{tavecchio18} used a simplified analytical treatment of the radiative cooling of the electrons. Here we adopt a self-consistent calculation of the electron energy distribution (i.e. numerical density of electrons with Lorentz factor in the range $\gamma-\gamma +d\gamma$ ) at different distances from the shock, $N(\gamma, d)$, by numerically solving the continuity equation including radiative losses, treated as in \cite{chiabghis99} using the robust fully implicit numerical scheme of \cite{ChangCooper}. 

In the shock rest frame emitting (and cooling) particles are advected downstream with velocity $v_{\rm adv}=\beta_{\rm d,sh}c=c/3$ (up to a distance $d_{\rm max}$, where we assume the emission is quenched by jet expansion/adiabatic losses, e.g. \citealt{MarscherGear85}). Therefore, we associated $t$ to a distance $d= v_{\rm adv} t$ from the shock. The distribution at time $t$ (and distance $v_{\rm adv} t$) is thus provided by the solution of:
\begin{equation}
\frac{\partial N(\gamma,t)}{\partial t} = \frac{\partial}{\partial\gamma} 
\left[ \dot\gamma_c(\gamma,t) N(\gamma,t)\right] + Q(\gamma) \delta(t),
\label{cont}
\end{equation}
where $\dot\gamma_c$ is the cooling rate, given by:
\begin{equation}
\dot\gamma _c= \frac{4}{3} \frac{\sigma_T c}{m_e c^2} [ U_B(d)+U_{\rm rad}] 
\gamma^2,
\label{eq:eed}
\end{equation}
where $\sigma_T$ is the Thomson cross section, $U_B(d)$ is the (total) magnetic field energy density at distance $d$ and $U_{\rm rad}$ is the energy density of the radiation field. For simplicity, in the calculations we assume that only synchrotron losses are relevant (i.e. we fix $U_{\rm rad}=0$) and we also neglect possible adiabatic and escape losses. We also neglect possible stochastic reacceleration/heating of electrons caused by interaction with turbulent fields in the post-shock flow (inferred to be of small intensity, \citealt{marscher22}).

We assume that relativistic electrons, instantaneously injected at the shock at $t=0$, follow a cut-offed power law energy distribution $Q(\gamma)\propto \gamma^{-n}\exp(-\gamma/\gamma_{\rm cut})$, with index $n$ and minimum and cut-off Lorentz factor $\gamma_{\rm min}$ and $\gamma_{\rm cut}$.  

In stationary conditions (suitable to model the quiescent state of the source during the observations analyzed here) the polarization can be calculated summing the Stokes parameters associated to each cell. In detail, Stokes parameters for each cell $i$ as a function of the frequency in the observer frame, $U_{\nu,i}$, $Q_{\nu,i}$ and $I_{\nu,i}$, are calculated following the standard formalism (\citealt{Lyutikov05,DelZanna06}). Finally, the total observed degree of polarization, $\Pi_{\nu}$, and the electron vector position angle (EVPA), $\chi_{\nu}$, are derived from the total Stokes parameters $U_{\nu}=\sum U_{\nu,i}$,  $Q_{\nu}=\sum Q_{\nu,i}$ and $I_{\nu}=\sum I_{\nu,i}$, by using the standard formulae:
\begin{equation}
    \Pi_{\nu} = \frac{\sqrt{Q_{\nu}^2+U_{\nu}^2}}{I_{\nu}}
\end{equation}
\begin{equation}
\cos 2\chi_{\nu}=\frac{Q_{\nu}}{\sqrt{Q_{\nu}^2+U_{\nu}^2}}, \;\; \sin 2\chi_{\nu}=\frac{U_{\nu}}{\sqrt{Q_{\nu}^2+U_{\nu}^2}}.
\end{equation}

\subsection{Application to 1ES 1101-232 and RGB J0710+591}

We reproduce the multiwavelength polarimetric data of the two sources adjusting the free parameters of the model. For both sources we fix the radius $r_j=4\times 10^{15}$ cm, the poloidal field intensity $B_z=0.03 $ G, the normalization of the self-generated field $B_{\perp, 0}=0.25$ G, the bulk Lorentz factor $\Gamma=20$. Electrons are injected with a slope $n=2.1$ with $\gamma_{\rm cut}=10^6$. The only different parameters for the two cases are the decay length and the slope of the self-produced field, $\lambda$ and $m$, and the viewing angle, $\theta_{\rm v}$, that we report in Tab. \ref{table:param}.

\begin{table}
\caption{Parameters of the models. \label{table:param}}
\centering
\begin{tabular}{cccc}
\hline
Source & $\lambda$ (cm) & $m$  & $\theta_{\rm v}$ (deg)\\
\hline
1ES 1101-232 & $5\times 10^{13}$ & 0.5 & 1.5\\
RGB J0710+591 &  $1\times 10^{12}$ & 0.2 & 1.25\\
\hline
\end{tabular}
\end{table}

\begin{figure}
\hspace{-0.5truecm}
\includegraphics[width=1.1\linewidth]{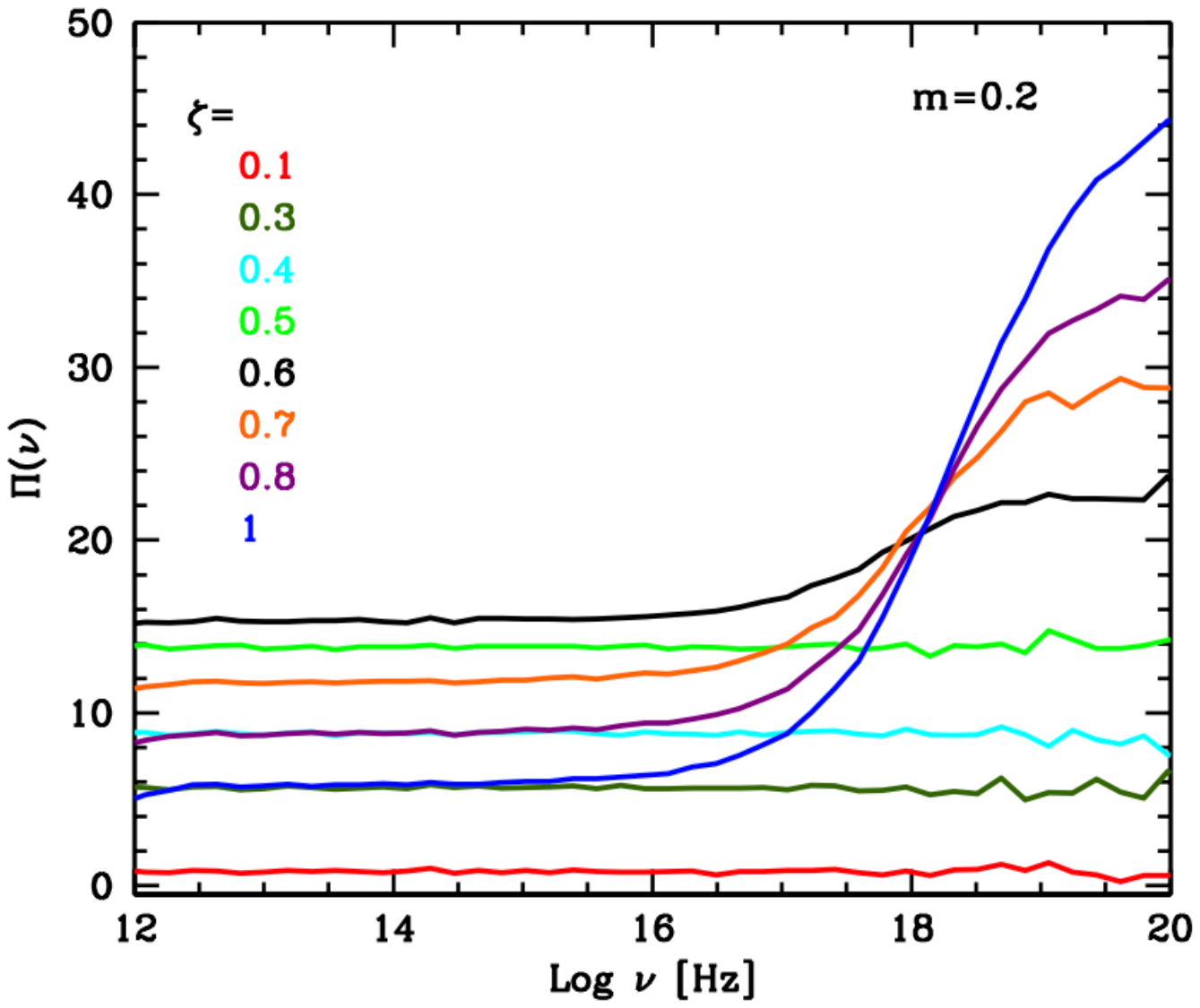}
\caption{Degree of polarization as a function of frequency for different values of $\zeta$ in the range 0--1, keeping fixed the value of the Doppler factor, for parameters of model 1.}
\label{fig:zeta}
\end{figure}

The main difference concerns the details of the profile of the self-generated field. In particular, the small $m$ used for RGB J0710+591 implies that the self generated field is important in the entire emission volume. With this configuration, when the jet is observed at small angle ($\theta_{\rm v}<1/\Gamma$) the degree of polarization is independent of the frequency. This is shown in Fig.\ref{fig:zeta}, which shows $\Pi(\nu)$ for different observing angles.

The critical quantity determining the observed (projected) structure of the magnetic field, and thus the resulting polarization, is the observing angle as measured in the jet frame, $\theta ^{\prime}_{\rm v}$. For small (observer frame) viewing angles, suitable to the sources considered here, it is useful to consider the parameter $\zeta =\Gamma\theta_{\rm v}$, the product of the bulk Lorentz factor of the emitting plasma and the observer-frame viewing angle. In particular, $\zeta=1$ (i.e. $\theta_{\rm v}=1/\Gamma$) corresponds to an observing angle in the downstream frame $\theta_{\rm v}^{\prime}=\pi/2$. In this situation, in the X-ray band one expects a rather large polarization, ideally close to the theoretical limit, since the region of emission is produced by the self-produced field whose projection is observed perfectly orthogonal to the jet axis. On the other hand, for decreasing values of the  jet-frame viewing angles, the observer has access to both projected (random distributed) $B_x$ and $B_y$ components and this results in a lower degree of polarization. In the limiting case $\zeta\rightarrow 0$ (which corresponds to $\theta ^{\prime}_{\rm v}\rightarrow 0$) the effective polarization approaches zero at all frequencies, since the observer is located exactly along the jet axis and sees an equal contribution of the two (randomly distributed) components of the self produced field, $B_x$ and $B_y$. Fig. \ref{fig:zeta}, reporting $\Pi_\nu$ for various values of $\zeta$ in the interval 0-1 clearly shows these effects. 

\begin{figure}
\hspace{-0.5truecm}
\includegraphics[width=1.05\linewidth]{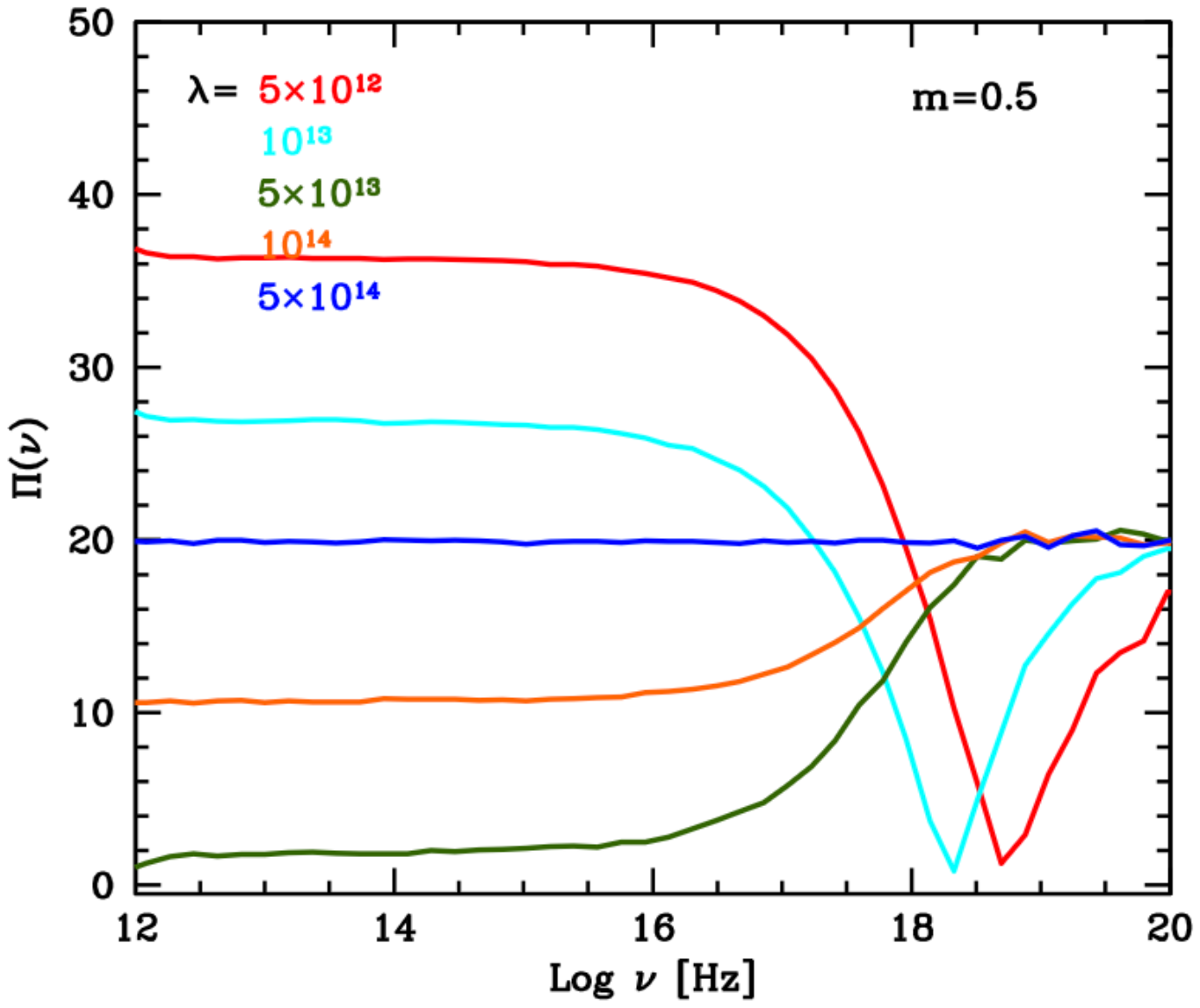}
\caption{Degree of polarization as a function of frequency for different values of $\lambda$, keeping fixed the value of the Doppler factor, for parameters of model 1.}
\label{fig:lambda}
\end{figure}

Fig. \ref{fig:lambda} instead illustrates the role of the decay length, $\lambda$, in shaping the degree of polarization (for the parameters used for RGB J0710+591). For very short $\lambda$  ($< 10^{13}$ cm) the self-genererated field decays very quickly and most of the low energy radiation (below the soft X-ray band) is produced in a volume where the field is dominated by the poloidal component, thus determining a high degree of polarization. For larger values of $\lambda$, the toroidal self-generated field is relevant in a large portion of the downstream volume, determining the decrease of the polarization fraction at lower frequencies (where both field components are important).

\end{appendix}

\end{document}